# Remote Teaching and Learning in Applied Engineering: A Post-Pandemic Perspective

**Mouhamed Abdulla and Weijing Ma**

School of Mechanical and Electrical Engineering, Faculty of Applied Science and Technology, Sheridan Institute of Technology, Toronto, Ontario, Canada
Email: {abdulmou, mawei} @ sheridanc.on.ca
Address: 7899 McLaughlin Road, Brampton, ON, Canada, L6Y 5H9

**ABSTRACT**

The COVID-19 pandemic significantly disrupted the educational sector. Faced with this life-threatening pandemic, educators had to swiftly pivot to an alternate form of course delivery without severely impacting the quality of the educational experience. Following the transition to online learning, educators had to grapple with a host of challenges. With interrupted face-to-face delivery, limited access to state-of-the-art labs, barriers with educational technologies, challenges of academic integrity, and obstacles with remote teamwork and student participation, creative solutions were urgently needed. In this chapter, we provide a rationale for a variety of course delivery models at different stages of the pandemic and highlight the approaches we took to overcome some of the pressing challenges of remote education. We also discuss how we ensured that hands-on learning remains an integral part of engineering curricula, and we argue that some of the applied changes during the pandemic will likely serve as a catalyst for modernizing education.

**INTRODUCTION**

The COVID-19 pandemic has had a devastating impact across the world, and overall, various sectors of society were impacted by this event. The educational sector was considerably affected by the pandemic as strict lockdown measures were imposed. Immediately, instructors and students were asked to proceed remotely with their teaching and learning. Despite this disruption, various disciplines that greatly depend on active hands-on learning had to find new and creative means to continue teaching in online mode.

Sheridan's engineering is one of these disciplines with program curricula that are contingent on learning-by-doing, system design, prototyping, and troubleshooting. In fact, these programs are designed with the objective of training students in the full lifecycle of real-world engineering based on conceptualization, design, implementation, and operation (CDIO) guidelines (Abdulla, Motamedi, & Majeed, 2019). At Sheridan, this educational CDIO framework is applied by particularly focusing on skills-based learning with experiential education, project-based education, work-integrated learning and applied research (Abdulla, Troop, & Majeed, 2020).

Unfortunately, with the reality of the pandemic, assembling, testing, and measuring engineering systems for hand-on CDIO learning and proof-of-concept was no longer a viable possibility. Lacking access to specialized laboratory facilities on-campus, faculty members and support staff had to promptly find solutions that would surmount this challenge, while guaranteeing a steady migration from face-to-face (F2F) to online delivery mode. Sheridan's course leaders in the different disciplines of electrical engineering (i.e., electronics, telecommunications, control systems, computer systems and power engineering) had to determine feasible workarounds and redraft related hands-on content for each of their respective courses.

While being cognizant that the impact and response to the global pandemic on education was unique in various parts of the world, in this chapter, we find it insightful to document and analyze pivotal events that impacted the higher educational sector in Canada, and particularly within the province of Ontario. We also outline disruptions encountered because of the prolonged shutdown of public spaces in the Greater Toronto Area, which is home to Sheridan's three campuses located in the municipalities of Mississauga, Brampton, and Oakville. In fact, the data suggests that Toronto is considered to have the longest continuous societal restrictions of any major city in the world (Levinson-King, 2021). Understanding these strict decisions taken over the span of the pandemic will help us better comprehend the rationale behind these rulings. We will also be able to identify any gaps in the evolving response from higher education to protect its learning community, while still ensuring quality education.

In this work, we also feature some of the effective and creative practices applied at Sheridan to overcome pressing educational obstacles as we pivoted to online and hybrid delivery. This includes, among others, topics related to interrupted F2F delivery, limited laboratory access, barriers with educational technologies, challenges of academic integrity, travel restrictions of international students, and obstacles with remote teamwork and classroom participation. By way of example, we also discuss how we swiftly adapted engineering courses for online education and explain the techniques used to support experiential learning with no or limited access to on-campus facilities. We reflect on the lessons learned and underscore the silver lining of the pandemic that catalyzed the gain of novel skill sets in virtual technology and educational innovation. Finally, we describe how these abilities are valuable for a post-pandemic future in transforming and modernizing the teaching of applied engineering.

**IMPACT OF THE PANDEMIC ON EDUCATION**

The response of the educational sector to the COVID-19 pandemic is generally related to regional trends. While Canada is the second largest country in the world, 87% of its 38 million inhabitants live in only the four major provinces of Ontario, Québec, British Columbia, and Alberta (Statistics Canada, 2021). Furthermore, in each of these provinces, most people cluster in and around the major cities of Toronto, Montréal, Vancouver, and Calgary. As such, the overall Canadian experience with the pandemic was unique with such geographically distant regions and cities scattered across the enormity of the Canadian land. Moreover, since healthcare is administered and managed at the provincial level, the individual response to the pandemic varied from one province to another. Due to unique regionality and jurisdictional responsibility, the impact of the pandemic on higher education varied accordingly.

To better understand the impact of the pandemic, consider the 7-day moving average of daily trend of new positive cases and fatalities of COVID-19 for the province of Ontario depicted in

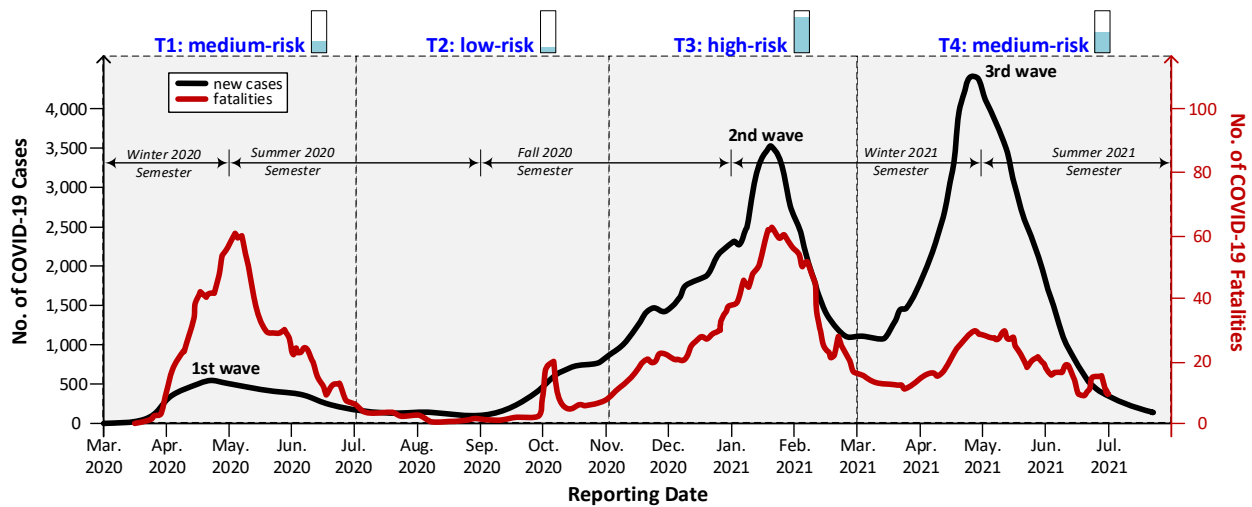

Figure 1.  Daily trend of COVID-19 cases for the province of Ontario, Canada.

Figure 1 using data from Public Health Agency of Canada (2021). This trendline is interesting to analyze not only because our campuses are in Ontario, but because it is the most inhabited province of the country with a share of roughly 39% of the population (Statistics Canada, 2021). Because of this size, it is no surprise that the shape of the normalized trendline for the pandemic in Ontario greatly resembles that of the country as a whole. Understanding the trend in Ontario can also give some indication of how the country is doing regarding this health crisis.

In Figure 1, the trend of new cases (shown in black) is plotted alongside the number of fatalities from the pandemic (shown in red). In hindsight, the coupling of these two plots will help us analyze the risk level over different duration intervals. In fact, we are looking at the impact of COVID-19 over the past 16 months of recorded data since the onset of the pandemic in March of 2020, and until the time of this writing in July of 2021. We can split the span of measurable data into four equally sized timeframes of four months each, which we denote respectively as T1, T2, T3 and T4 in Figure 1. For context to the educational sector's response, we also overlay in this figure the five semesters (i.e., Winter 2020, Summer 2020, Fall 2020, Winter 2021 and Summer 2021) that were impacted since the start of the pandemic and until now.

With this arrangement, we can assess the risk level at each interval of four months by multiplying the likelihood of an event (i.e., low, or high probability of positive COVID-19 cases) with the severity of the event (i.e., low, or high consequences of the coronavirus, where low impact is associated with recovery and high impact with fatality). As shown in Figure 1, the pandemic trendline in Ontario (also in Canada) indicates three obvious waves of COVID-19 cases, where the first wave occurred in T1, the second wave in T3, and the third wave in T4.

As presented in Table 1, this information paired with the fatalities curve results in a coherent and simple two-by-two risk matrix for the different time intervals. As shown, a rough estimate of the risk suggests a low-risk event in T2, a medium-risk event in T1 and T4, and a high-risk event in T3. Due to an incredible amount of uncertainty, erring on the side of safety and taking "extraordinary measure in an abundance of caution" (Clay, 2020) at the start of the pandemic by Sheridan was absolutely understandable. However, in retrospect, it could be argued that a data-driven approach for risk assessment could have perhaps yielded a more proportionate response by the provincial government and the educational sector for general operations. This could have lessened the complexity of migration by giving more time for online transition.

Table 1. Risk matrix of COVID-19 pandemic on higher education in Ontario, Canada.

| Risk = Prob. × Impact | High Impact | | Low Impact | |
|---|---|---|---|---|
| Low Probability | Period: Mar. → Jul. 2020<br>Semesters: Winter/Summer 2020<br>Risk Level: Medium<br>Implemented Mode: Remote Education | T1 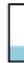 | Period: Jul. → Nov. 2020<br>Semesters: Summer/Fall 2020<br>Risk Level: Low<br>Implemented Mode: Hybrid/Remote Education | T2 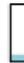 |
| High Probability | Period: Nov. 2020 → Mar. 2021<br>Semesters: Fall 2020/Winter 2021<br>Risk Level: High<br>Implemented Mode: Hybrid/Remote Education | T3 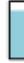 | Period: Mar. → Jul. 2021<br>Semesters: Winter/Summer 2021<br>Risk Level: Medium<br>Implemented Mode: Hybrid/Remote Education | T4 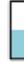 |

Certainly, in the midst of a life-threatening pandemic, it is generally impossible to predict with high accuracy and high precision the extent of an infectious disease wave and assess, *a priori*, its risk level. This is an example of a known-unknown event, as we know the existence of the pandemic, but lack full information and understanding of the risk level at different stages and time intervals. Experts on pandemics could use mathematical models to forecast the spread of infectious disease. But the fact remains that these are likelihood predictions based on available data and input assumptions, which sometimes result in over- or under-estimation of the risk. As the statistician George Box is attributed to have said, "all models are wrong, but some are useful". Nonetheless, the situation is gradually changing as infectious disease models are improving in sophistication and accuracy (Strain, 2020).

**ADAPTIVE COURSE DELIVERY MODELS**

As the prediction capability of infectious disease improve, government agencies could for instance assess the risk using highly reliable and advanced analytical tools and offer recommendations and guidelines to the educational sector for a calculated and adequate response for permissible operations. For example, depending on the estimated risk level, a specific course delivery mode can be selected from the spectrum of available choices outlined in Figure 2. On one end of the spectrum, under a high-risk situation with curfews and maximum restrictions, the course delivery should exclusively be in virtual mode with no F2F component. On the other end of the educational spectrum, when no imminent risk exists or is known, traditional on-campus education is naturally applied with the option of including some virtual delivery component (e.g., implemented with a flipped classroom model). Between these two extremities, a flexible or hybrid education mode can be considered where the former can be applied for low-risk, and the latter for a medium-risk event.

To be precise, in the flexible education model, students are given the choice to either attend classroom activities or be present remotely using educational platforms and learning management systems (LMS). This model can also be compatible with the flipped classroom approach (Bishop & Verleger, 2013), where the learning and content dissemination is achieved outside class time, say through pre-recorded video lectures, and the application and practice is done in F2F fashion on-campus and/or synchronously with livestreaming sessions. This approach effectively uses class time for skills-based learning, learning-by-doing, experiential learning, problem-solving strategies, interactive discussions, and brainstorming sessions for major deliverables. Indeed, in comparison to traditional lecturing, the flipped classroom model provides greater opportunity for collaborative learning during class time (Jdaitawi, 2019).

On the other hand, the hybrid education model is different from flexible learning as students are not really given the choice to attend campus for F2F delivery or join class virtually. For applied learning disciplines such as engineering, the hybrid model can be used primarily to accommodate hands-on labs that are impossible or impractical to conduct off-campus through simulations, augmented and virtual reality (AR/VR), affordable off-the-shelf equipment or using development kits. Under this special circumstance, authorized students are permitted campus entry, provided they follow personal protective equipment guidelines from the school, the local authority, and the Ministry of Health. For the coronavirus pandemic, this means that students, support staff and faculty should all wear medical facemasks, safety glasses, use hand sanitizer upon entry to the lab, and maintain physical distancing of at least two meters.

To further minimize the risk of spreading the pandemic under the hybrid model, lectures are exclusively delivered online. In accordance with pre-established school guidelines, the instructor can be given the choice to either deliver content through pre-recorded videos or present a lecture in real-time streaming using LMS applications or teleconferencing software. It is worth noting that there are unique advantages with both synchronous and asynchronous course delivery. In a live session, teaching and learning is more interactive and engaging as students' questions, clarifications and instructor feedback can all be done immediately in real-time. Meanwhile, in a recorded lecture, students have greater flexibility to move at their own pace and bypass mastered topics through video timestamps, they can learn at any day and time of their choice and the learning can revolve around their daily schedule. Yet, asynchronous teaching is often more time consuming than synchronous delivery. In essence, it is a form of rehearsed performance with multiple iterations so as to ensure a final recorded content that is technically accurate, professionally produced and of high quality.

Informed by various classroom feedback, many of our engineering students preferred the synchronous delivery as a motivator to their scholastic success. Moreover, during various COVID-19 lockdowns, where stay-at-home orders were enforced, any possibility of social connectivity, albeit remotely, brought a degree of contentment and a sense of togetherness with the students, their peers, and the class instructor. As indicated by Pietrabissa and Simpson (2020), isolation associated with this pandemic has resulted with various psychological consequences, such as, loneliness, depression, and anxiety. Recognizing that some students may have felt isolated, it was important to show compassion, consideration, and a sense of community during this difficult period.

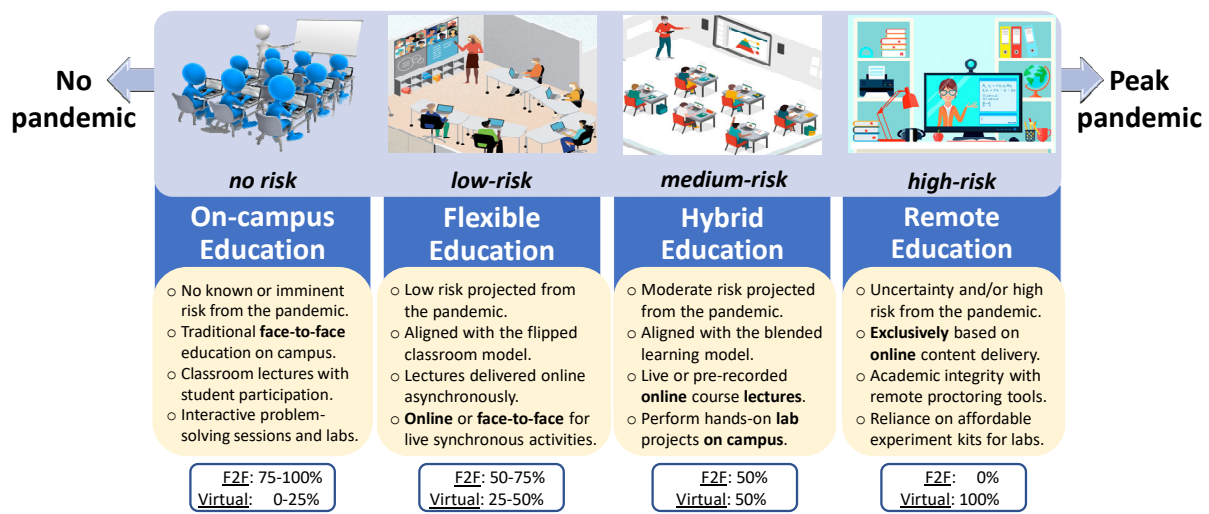

Figure 2. Variety of course delivery models at different stages of the pandemic with recommended split between F2F and virtual delivery mode for each option.

Moreover, in addition to the live online classes, many students expressed their interest in obtaining access to the recording of synchronous sessions for review before exams. With different LMS applications (e.g., Brightspace or Blackboard), it is generally possible and easy to automatically record the live sessions and publish them for students' access. Meanwhile, there are other learners that favored a more curated pre-recorded video lectures that only focused on the content without the excess of in-class interactions (i.e., without questions, comments, classroom interruptions, etc). Accommodating the different learning preferences of all students certainly increases the instructor's workload beyond what is expected. Judgment should be made to proceed with the best course of action for one's specific classroom.

Alternatively, an instructor can apply a blended learning approach (Hill, 2012). For example, the course can be designed with a mixture of pre-recorded video lectures for content delivery (e.g., virtual/asynchronous: 25%); live online streaming for learning-by-doing exercises, problem-solving strategies, and interactive discussions (e.g., virtual/synchronous: 25%); and on-campus access for hands-on labs (e.g., F2F: 50%). Of course, any other derivatives of this setup that is compatible with learning outcomes of a course is possible to apply for the hybrid educational model with blended learning.

At present, we can circle back to Figure 2 and observe the recommended split proposed between F2F and virtual delivery for the different educational models. Notice that the importance of online delivery increases as the risk varies from low to high, and as we move from F2F to remote education. Conversely, as the risk level subsides with time, the role of on-campus learning increases. The COVID-19 pandemic showed us that although being physically present on-campus is important on so many levels (e.g., connectedness and access), if done well, remote education can function quite effectively. Therefore, in a post-pandemic world, when we return to a *new normal*, the convenience of virtual delivery with hybrid or flexible modes will likely still be part of several courses and programs. Irrespective of risk levels, the use of blended and flipped classroom approaches with a combination of on-campus, synchronous and asynchronous delivery can provide greater variety in learning. Moreover, these approaches will make the educational sector more accessible for students that work, have family obligations, health and mobility concerns, transportation challenges, etc.

Table 2. Key educational challenges faced during the COVID-19 pandemic.

| Main Challenges | Stakeholders in Education | | | |
|---|---|---|---|---|
| | Students | Faculty | Staff | Admin. |
| **Interrupted F2F Delivery**: Redesign course curricula and digitize all pertinent learning contents and evaluations for remote delivery. | | ✓ | ✓ | |
| **Limited Laboratory Access**: Creatively think of new experiential hands-on exercises with limited or no access to on-campus equipment. | | ✓ | ✓ | ✓ |
| **Barriers with Educational Technologies**: Train instructors the full capacity of e-learning platforms for virtual classroom delivery. | | ✓ | | |
| **Concerns with Academic Integrity**: Uphold academic integrity and safeguard against plagiarism using anti-cheating tools and strategies. | | ✓ | | ✓ |
| **Travel Restrictions**: Inability of international students to travel for synchronous learning with F2F or hybrid course delivery mode. | ✓ | ✓ | ✓ | ✓ |
| **Truncated Semesters**: Regular semesters were shorten by a week, where the extra time was allocated for additional course preparation. | ✓ | ✓ | ✓ | |
| **Obstacles with Remote Participation**: Hesitation of some students to actively participate (chat, audio, and video) in synchronous remote delivery. | ✓ | ✓ | | |

**PEDAGOGICAL CHALLENGES**

Over the past year and a half of teaching under the risk of COVID-19, we faced various educational hurdles. We compiled a list of the most prominent challenges in our teaching practice in Table 2. As outlined, some of these difficulties are more generic for education, while others are discipline specific. The table is also organized as a function of the different stakeholders in education that are most impacted by each obstacle (i.e., students, faculty members, technical and support staff, and school administrators). Below, we briefly explain some of our responses to these challenges:

- *Interrupted F2F delivery*: To overcome this challenge, we had to redesign and digitize all pertinent components of engineering courses, including, lectures, labs, project deliverables, assignments, quizzes, and exams. Overall, this was a very laborious task, especially in early days of the pandemic while instructors were gaining new skill set with diverse LMS tools for online delivery.

- *Laboratory access*: Experiential engineering learning was accomplished using different approaches over remote delivery. Course leaders and faculty members had to redraft experimental setups and handouts by amending labs with major reliance with on-campus laboratory equipment. Quite often, new labs were designed to fit the reality of remote education. These lab exercises focused more on circuit- and system-level simulations, computer programming, AR/VR platforms, IoT-based development kits, circuit components and off-the-shelf measurement equipment.

- *Technological barriers*: Faculty members had to be trained and retrained with LMS tools that are exclusively necessary for remote education, including, features for synchronous and asynchronous teaching, online examinations, and remote teamwork. The role of Sheridan's Centre for Teaching and Learning was instrumental in assisting faculty members improve in proficiency with new skill sets for online education.

- *Academic integrity*: Students were advised at the start of each semester that we will use remote e-proctoring software with live monitoring of audio, video, and screen activity for the purpose of upholding academic integrity. We also used tools that automatically detect and rate the likelihood of cheating behavior using facial detection technology. For written deliverables, we used Turnitin, an online plagiarism detection service applied to verify the originality of students work.

- *Travel restrictions*: Roughly 20% of Sheridan's students are international with over 100 countries represented. Some of our international students were restricted from entering the country. Due to time zone differences, livestreaming of lectures and synchronous examination was always a challenge. Teaching was bifurcated between domestic and international students as deadlines were usually extended for international students due to time zone differences, delays in accessing remote lab equipment, intercontinental travel, and mandatory quarantine requirements.

- *Truncated semesters*: Since the onset of the pandemic, five consecutive semesters operated with 13 weeks rather than the usual 14 weeks. The shortened week in each semester was given to instructors to help them in migrating content from F2F to remote delivery. Despite the gain in preparation time, course leaders had to redesign class plans to accommodate all learning objectives be met within the readjusted period.

- *Remote participation*: Although students had the option to use chat, microphone, and camera features for real-time interactions during remote sessions, yet, some learners were still reluctant to participate with questions and comments. To encourage active online participation, we modified the evaluation plan of third year wireless communication courses by allocating a 10% project component. For the deliverable of this component, students had to synchronously present to the entire classroom their research findings in an oral camera presentation with slides. In weeks following this learning activity, we observed a noticeable improvement in remote participation.

**IMPLEMENTED PRACTICES**

Over the span of the pandemic, we implemented different educational practices. Upon reflection, we can attribute some of our engineering schools' accomplishments to the following key factors:

***Curriculum Transition in Phases***

When the entire province of Ontario went into lockdown, the Winter 2020 semester was well underway with only four weeks of teaching remaining. By that time, most students have gained hands-on skills and equitable knowledge of lab equipment. This knowhow assisted students' comprehension of new and advanced materials covered virtually during the remaining parts of the semester. As the duration of the pandemic extended, and subsequent semesters were switched to online or hybrid mode, returning students during these semesters benefited significantly from the experience gained in the initial transition. This unintentional phased approach has been an important contributor to the successful delivery of our programs.

***Ready-to-deploy Hardware and Software Tools***

To serve the different learning needs and styles of our students, we embraced the concept of universal design (UD) with our teaching and learning practices (Smith, 2012). The UD approach provides a list of principles in developing course instructions and materials (Burgstahler, 2009). For instance, we used a suite of pedagogical tools to teach engineering and technology concepts with hands-on exercises. This included the use of LMS applications, software packages, simulation techniques and hardware equipment. Meanwhile, we also ensure that we had the necessary number of licenses for the various lab applications. This allowed our students to begin their remote learning without excess delays.

Besides, online platforms for the organization of course materials was employed by instructors and learners. Even before the pandemic, LMS applications have been around and widely utilized in higher education (Katz, 2013; Kuran, Pedersen, & Elsner, 2017). Platforms such as Brightspace, Blackboard, Canvas and Moodle are some of the widely used virtual learning environments in Canada (Peters, 2021). Our school has been consistently using LMS applications to organize and curate course materials. During the initial transition from F2F to online delivery, LMS features that were previously underutilized were proven extremely helpful.

Throughout the pandemic, our virtual learning environment provider regularly updated the new online examination tools to include different features for students, including among others, identification and verification, online-invigilation, library access, course-based chatrooms and forums, video homework, and project spaces. These updated features have significantly expanded the capacity of the virtual learning platform, making it a comprehensive tool for course delivery and management of evaluations and deliverables. Overall, the experience with various software and hardware technologies facilitated our school's seamless transition into the alternate delivery mode.

### *Consolidated Effort and Response*

Another key reason for the effective transition to remote learning can be attributed to our school leaders and administrators. They were instrumental in entrusting subject matter experts in devising a transition plan. The goal was to ensure a feasible and consistent plan with minimal impact on the learning outcomes of courses. Empowered by administrative support, faculty members created contingency plans for all engineering courses within our programs.

For instance, our faculty members, support staff and technologists created an equipment sign-off procedure to provide senior students access to portable equipment for remote hands-on learning activities. As for first-term courses, they were switched to a fully online format with video lectures, simulations, remote labs, and online evaluations. Meanwhile, in engineering and technology education there were two type of courses that caused additional challenges: (i) courses that are reliant on expensive equipment, where equipment sign-off is not feasible; and (ii) courses that contain significant hands-on training, where simulation alone was insufficient for meeting the learning objectives. For the latter, some instructors developed VR-based laboratory exercises. This approach provided a unique solution to compensate for the lack of practice with actual on-campus equipment.

### *School's Agile Response*

Since the success of the initial transition to remote education, Ontario has gone through various waves of COVID-19 cases followed by lockdowns (see Figure 1). During this time, some of our community members were infected or exposed, and required to quarantine for an extended period. We made sure that all our livestreaming lectures be recorded so that impacted students that missed lectures can review the content asynchronously. When students are in quarantine due to either case confirmation or possible exposure, we offered accommodation for laboratory deliverables.

Overall, compared to previous academic years, our programs have sustained and produced a similar retention rate during the pandemic. Moreover, informed by students' key performance indicator survey, on-campus education pre-pandemic and remote education during the pandemic resulted in a comparable level of satisfaction. These encouraging feedbacks and observations are driving us to further improve our remote and hybrid course delivery.

## LESSONS LEARNED AND STRATEGIES FOR THE FUTURE

Through this unprecedented pandemic, the Ontario higher educational sector has gone through a tremendous amount of pressure to maintain quality education while ensuring the safety of its learning community. Via creative problem-solving and agility, we were able to overcome critical educational challenges. Reflecting on what has been done since the start of the pandemic, we feel the following key points resonate with the majority of our experiences and can serve as lessons and strategies for the future:

*Transition into Smaller Class Sizes*

As a consequence of physical distancing on-campus due to the pandemic, we noticed some of the unique benefits of smaller class sizes (i.e., less students per classroom). While smaller class sizes have been an indicator of scholastic quality in higher education (Wright, Bergom, & Bartholomew, 2019), engineering schools have typically preferred large lecture halls with hundreds of students. Although teaching F2F to many students at once is cost-effective, the educational benefits of lecturing to a smaller audience cannot be underestimated. Some of the key benefits of smaller classes for engineering education are:

- *Enhanced student engagement*: Smaller classes often lead to a more intimate learning environment. It reduces the sense of anonymity in the classroom, and it enhances student-teacher interactions. Of course, student's voice is amplified in such an environment as many students feel less reluctant to ask questions and provide comments. In addition, in a smaller class, student concerns can be addressed faster by the instructor. This leads to a more noticeable impact of their contribution and engagement in the learning environment.

- *Effective student-teacher interaction*: In a small class, more attention is given to individual student's learning progress and hands-on practice. For example, this is vital during the prototyping phase, where design, implementation, and troubleshooting skills require close consultation with the instructor. Combined with other technical and curricular improvements, a smaller and personalized class can provide a positive impact on students' learning.

- *Improving experiential learning*: Since engineering education includes the training of students' hands-on skills beyond theoretical fundamentals, engineering laboratories are a critical component of an effective curriculum (Feisel & Rosa, 2005). Smaller classes are particularly suitable when teaching engineering and technical content, where hands-on practice and instructor feedback are essential.

With the modification of learning spaces for safety purposes due to the pandemic, our students appreciated the inherent benefits of smaller class sizes. Post-pandemic, smaller classes with more personalized teaching approaches need to be emphasized and prioritized for effective on-campus engineering education.

*Maintain a Balance between Online and On-Campus Education*

While it is widely recognized that smaller classrooms are beneficial to learning, the debate between online, hybrid and on-campus delivery has been ongoing in literature (Beynon, 2007). During the pandemic, remote and hybrid learning became a necessity instead of an option. This rapid paradigm shift posed enormous challenges to instructors, of which many were unfamiliar with the subtleties of online education. At the same time, students were forced to switch their entire education remotely, and overnight, they had to change their usual learning practices. Students as well encountered various difficulties with online learning. With time, both students and faculty members overcame major obstacles and, to some extent, embraced some of the flexibilities and benefits of remote education.

However, embracing online teaching in the interim should not necessarily be an indication that programs should switch to long-term remote education. Instead, extensive research and investigation is needed on the effectiveness of the different course delivery models. Certainly, an abundance of experiences have been gained on delivery models during the course of the pandemic. In particular, the topic of virtual and hybrid education has generated an increasing interest during the pandemic (Dhawan, 2020; Rapanta, Botturi, Goodyear, Guàrdia, & Koole, 2020). These research results, and the like, should be assessed carefully by different stakeholders in education, and through a cost-benefit analysis, a decision can be made on the best approach to follow for each specific discipline.

Meanwhile, for engineering education, online teaching and learning is not a panacea post-pandemic. For various engineering specializations, on-campus hands-on education is paramount. Thus, an exclusive online delivery may not necessarily be a sustainable solution for all applied disciplines. As discussed earlier, applying a blended or a flipped classroom approach with a mix of online and on-campus components will likely produce a better educational alternative.

For such a blended learning transition, designing effective online and on-campus learning modules by relying on current expertise of faculty members may not be sufficient. Rather, it will also depend on the continual professional development of instructors. For such reasons, institutional support, funding and research opportunities in the scholarship of teaching and learning should be available post-pandemic.

### *Leverage Multiple Digital Learning Methods*

Over the duration of the pandemic, online education is implemented using different technological tools and methods. Today, many course modules are developed using digital learning techniques that include, e-learning platforms, courseware programs, video conferencing, video recording, voiceover slide presentation, computer-based simulation, remote examination, etc. With the rise of digital learning methods, educators and researchers are now interested in exploring how these technologies complement each other, so as to enhance the learning experience of students. With growing awareness and competence of digital learning skills by faculty members, the integration of these methods to support students learning will be an ongoing trend for the future of higher education.

Besides, with the rapid introduction of alternative delivery modes during the pandemic, engineering students have been exposed to various ways of remote and hybrid education. Within a relatively short period, students were able to experience and compare the effectiveness of learning under different delivery models. Based on various accounts, this experience shed light on their learning method of choice. Post-pandemic, instructors should still strive at exposing their students to multiple digital learning methods and delivery modes, in an attempt at aiding their students discover their learning preferences.

### *Balance in Equipment and Technology Investment*

In our electrical engineering laboratory facilities, we have various industry-grade equipment for measuring and testing electrical systems and signals. During strict lockdowns, our on-campus lab was inaccessible, and therefore unusable. With remote delivery, these on-campus equipment that are not connected to the internet were significantly underutilized.

Post-pandemic, we recommend a balanced approach of investing with hardware equipment, software applications, and cloud-based technology useful for remote and on-campus course delivery. The planning, procurement and curricular design of future laboratories should for instance focus on providing:

- *Portable systems*: Despite being small, portable, and generally less powerful than on-campus equipment, these devices (e.g., off-the-shelf equipment, development kits, IoT devices, SDR units, etc.), provide an affordable alternative to state-of-the-art devices. They do not require specialized facilities for operations, and they can be transported for remote education by students. Students can gain confidence and entry-level skills by using, testing and measuring with such devices.

- *Remote instrument access*: Setting up network and cloud-based remote access systems can be used to connect remote learners to centralized on-campus equipment. This sophisticated setup provides online accessibility to state-of-the-art equipment that are physically located in different laboratory facilities on-campus. This requires proper online platforms and compatible devices to make such equipment remotely accessible.

Thus, in addition to traditional on-campus equipment, portable devices and remote instrument access will all be necessary for hands-on engineering education for future course delivery with a blended learning model. Overall, the experience of the COVID-19 pandemic should be a reminder that we need to smartly plan laboratory environments that can be compatible with the different course delivery models.

**CONCLUSION**

As the story of COVID-19 and its more transmissible variants is still unfolding, we still lack the full understanding of when the educational sector will return to a full *normal* state. With the increase of vaccinated individuals and the prospect of reaching herd immunity, a *new normal* seems closer within reach. With this in mind, we aimed in this chapter to provide a rationale for the educational response over the past year and a half of teaching under the impact of the pandemic in Ontario, Canada. For us, this unique experience was instrumental and a paradigm shift, as it propelled us to significantly advance with educational innovation during the pandemic. With this momentum, the trend of further modernizing engineering education will continue post-pandemic, as there are way more technologies and novel practices that we are interested to experiment and apply in our school.

Overall, we see great challenges and opportunities that require special attention from the higher educational system. For instance, engineering disciplines that rely heavily on experiential learning need to explore new ways of course delivery when in-person training is reduced. With greater reliance on online and remote delivery, technological innovations will provide alternative ways for in-person access to state-of-the-art facilities. Meanwhile, upholding academic integrity and deterring plagiarism with remote evaluations remains a constant struggle. Stakeholders in education will need to creatively think of how to engage and evaluate student performance for an adaptive course delivery model approach.

Despite some of these challenges, remote delivery has proven effective thus far. However, for the future of engineering education post-pandemic, we believe that a *blended learning* approach with a combination of *online* and *on-campus* delivery with *synchronous* and *asynchronous* learning contents can provide a much-needed variety in education. This change will undoubtedly make the higher educational sector more flexible and accessible to students.

On the whole, the discussions and discoveries over the course of the past few months since the onset of the pandemic will certainly serve as a springboard and a catalyst for the future of applied education.

**ACKNOWLEDGEMENTS**

The authors would like to acknowledge the leadership, staff, and faculty members from the School of Mechanical and Electrical Engineering, and the Centre for Teaching and Learning at Sheridan Institute of Technology for various discussions over the course of the COVID-19 pandemic. Their thoughtful feedback, valuable input, continuous support, and encouragement in the scholarship of teaching and learning led in part to the compilation of this work.

**BIOGRAPHICAL INFORMATION**

***Mouhamed Abdulla***, Ph.D., is a Professor of Electrical Engineering at Sheridan. Since 2015, he was a Marie-Curie Fellow at Chalmers University of Technology, and a Visiting Fellow at Tsinghua University. Before that, he was an NSERC Postdoctoral Fellow at University of Québec. He also has 7 years of industrial experience with IBM Canada Ltd. He received respectively a B.Eng. (with Distinction) in Electrical Engineering, an M.Eng. in Aerospace Engineering, and a Ph.D. in Electrical Engineering, all at Concordia University in Montréal.

***Weijing Ma***, Ph.D. is a Professor with the Faculty of Applied Science and Technology at Sheridan College in Toronto, Ontario, Canada. She is a registered professional engineer in Ontario, Canada. Weijing received her Ph.D. degree in Electrical and Computer Engineering from the University of Toronto under the supervision of the late Professor Michael L. G. Joy and Professor Adrian Nachman. Her research interests include medical imaging, computer vision, and biomedical IoT applications.